\ifdefined\pdfminorversion\pdfminorversion=7\relax\fi
\documentclass[conference,compsoc]{IEEEtran}

\usepackage[nocompress]{cite}
\usepackage{amsmath,amssymb}
\usepackage{graphicx}
\usepackage{booktabs}
\usepackage{array}
\usepackage{tcolorbox}
\usepackage{url}
\usepackage{tabularx}   
\newcolumntype{Y}[1]{%
  >{\hsize=#1\hsize\linewidth=\hsize
    \raggedright\arraybackslash}X%
}
\newtcolorbox{researchquestions}[1]{
  title={#1}, colframe=black!20, colback=black!2.5,
  coltitle=black, fonttitle=\bfseries, boxrule=0.5mm, arc=1mm,
  before skip=\smallskipamount, after skip=\smallskipamount
}
\newcommand{\vt}{VisitTrace}
\newcommand{\takeaway}[2]{\par\noindent\textit{\textbf{Takeaway #1.}} #2\par}
\newcommand{\guideline}[2]{\par\noindent\textbf{\textit{Guideline #1.}} #2\par}

\ifdefined\SingleColumnOverview
  \newenvironment{overviewfigure}{\begin{figure}[b]}{\end{figure}}
\else
  \newenvironment{overviewfigure}{\begin{figure*}[t]}{\end{figure*}}
\fi
\newlength{\lifecyclewidth}
\newcolumntype{L}[1]{>{\raggedright\arraybackslash}p{#1}}

\begin{document}
\title{SoK: How Are Flows Labeled? A Lifecycle Audit of Encrypted Traffic Classification Benchmarks}
\author{\IEEEauthorblockN{Sizhe Huang, Shujie Yang}}
\maketitle

\begin{abstract}
The performance of encrypted traffic classification (ETC) methods is typically evaluated on public benchmarks. Ideally, benchmark scores should reflect recognition of target traffic in real-world settings. Yet unsupported flow labels, excluded target flows, and insufficient evidence for verification can steer model training and evaluation away from that task.
We audit 13 public ETC datasets across collection, labeling, and downstream use, identifying two recurring problems. \textit{Coarse inheritance} assigns collection-level labels to individual flows without corresponding attribution evidence; \textit{overstrict filtering} excludes target flows that fail network-identifier matching rules. Analyses of public datasets and controlled experiments using VisitTrace, our independently attributed website-traffic corpus, show that these practices can conceal poor recognition of intended target flows behind higher benchmark scores. Including previously excluded target flows in VisitTrace evaluation reduces balanced accuracy by up to 42.87 percentage points. These distortions can persist through dataset reuse, misleading judgments about method improvements and steering research toward benchmark-specific requirements. We propose \textit{correctness, representativeness, and auditability} as fundamental requirements for reliable ETC benchmarking.
\end{abstract}

\section{Introduction}
\label{sec:introduction}

Encrypted traffic classification (ETC) aims to identify applications, services, and user activities from encrypted network traffic. Existing research widely uses public datasets to train and evaluate ETC methods as encryption protocols progressively restrict the information available to observers. Public benchmarks have become essential research infrastructure. Judgments about method improvements and technical progress increasingly depend on the results these benchmarks provide~\cite{sok2025,sugar2025,etbert2022,raji2021benchmark}.

\begin{figure}[b]
\centering
\includegraphics[width=0.9\linewidth]{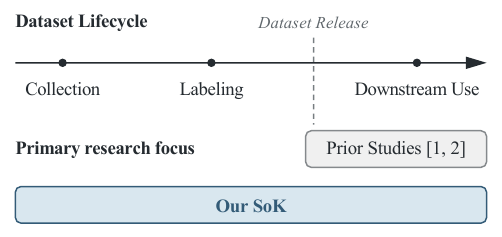}
\caption{Scope of this audit and prior studies across the ETC dataset lifecycle.}
\label{fig:research-scope}
\end{figure}
 
Previous studies have primarily focused on the post-release stages of ETC benchmarking (Fig.~\ref{fig:research-scope}). They analyze how explicit identifiers, connection-specific fields, data splits, and protocol filtering affect classification scores, and recommend corresponding processing practices~\cite{sok2025,sugar2025}. These studies establish the influence of input features and evaluation settings. 

However, \textbf{evaluation validity is already affected during dataset construction~\cite{engelen2021,holland2026pcapml,tsipras2020}.}
A traffic capture may contain both target-related connections and unrelated background traffic. Constructing an ETC dataset therefore requires deciding which flows are retained and how each flow is labeled\cite{soukup2021quality}. 
 These decisions determine whether benchmark scores reflect a model's ability to recognize target flows in the intended classification task, and thus affect judgments about progress in ETC. Verifying them requires corresponding evidence from collection and labeling, which released datasets may no longer retain.

We conduct a lifecycle audit of 13 public ETC datasets, tracing how decisions made during collection and labeling carry into downstream use. We identify two key construction problems: \textbf{coarse inheritance} and \textbf{overstrict filtering}. Fig.~\ref{fig:construction-problems} illustrates how these problems arise when captured traffic is converted into labeled samples.

\begin{overviewfigure}
\centering
\includegraphics[width=0.9\linewidth]{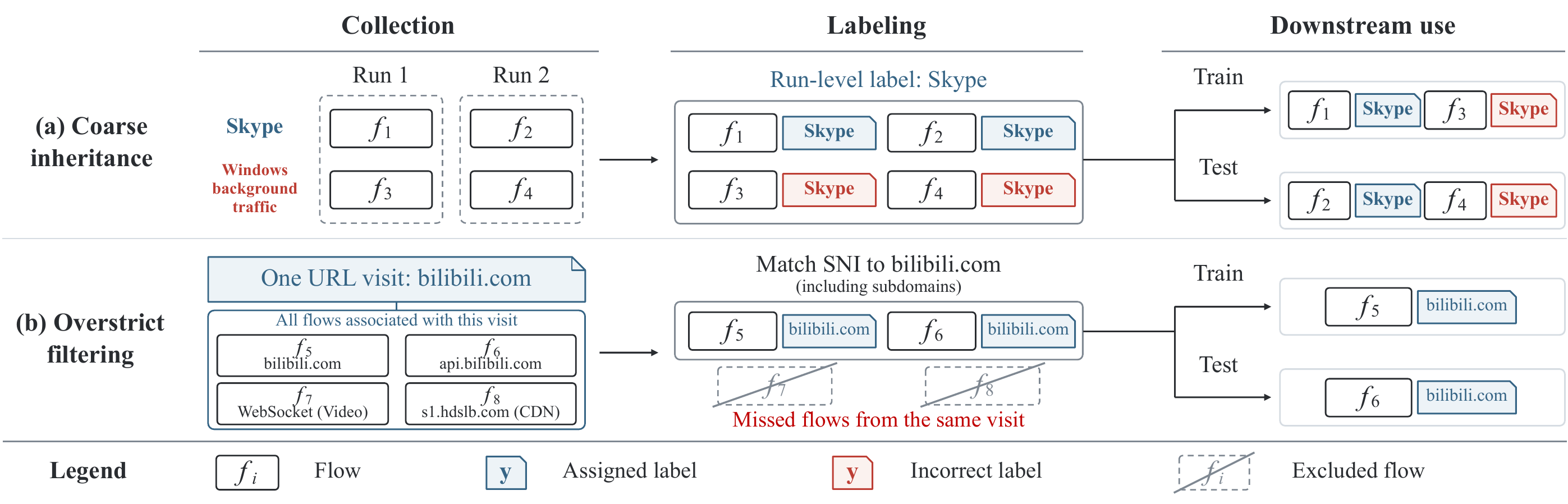}
\caption{(a) Windows background flows inherit the Skype label. (b) An SNI filter excludes two flows from the same website visit.}
\label{fig:construction-problems}
\end{overviewfigure}

\textbf{Coarse inheritance.} Dataset builders record evidence at the level of a collection activity, but treat it as evidence of flow-level identity. They then assign the application or website label to every flow collected during the activity~\cite{etbert2022,iscxvpn2016}. Yet an activity can include flows generated by sources other than the target application. A model may therefore learn to predict the collection activity rather than the claimed application identity, calling the correctness of the labels into question.

\textbf{Overstrict filtering.} Dataset builders generate labels from network identifiers carried by individual connections and retain only flows that the labeling rule can identify. Labeling connections using identifiers such as Server Name Indication (SNI)~\cite{etbert2022,sok2025} overlooks the fact that a website/application visit may retrieve resources over multiple flows and involve different forms of encryption, including TLS and end-to-end encryption (E2EE). If inclusion requires SNI to match the target domain, only TLS flows with matching names remain. Other connections belonging to the visit are excluded. Scores then describe recognition of the retained flows and do not adequately represent the intended classification task.

Our experiments quantify the effects of coarse inheritance. In ISCXVPN, after constructing features based on prior audits~\cite{sok2025,sugar2025}, 69.99\% of flows share identical inputs with flows assigned other application labels. A model cannot classify all such flows correctly, and the empirical balanced-accuracy (BA) ceiling is only 84.38\%. Under coarse inheritance, some classification requirements are therefore impossible to satisfy with the specified input. We further investigate this problem using MIRAGE, which provides evidence to establish accurate application labels for individual flows. In a controlled majority assignment used to examine coarse inheritance, replacing each flow's label with the most common application among the other matched flows in its capture changes only 4.54\% of the labels, yet removes 23 of the original 47 application classes from the training labels.

To examine overstrict filtering, we construct \textbf{VisitTrace}, a corpus of traffic captured during real website visits. We correlate HTTP Archive (HAR) requests with Chrome
DevTools Protocol (CDP) request events, then link them
to Chromium NetLog connections and captured flows
to establish visit-to-flow attribution independently of SNI. Applying the rule that retains only flows whose SNI matches the visited website's domain leaves just 13.58\% of the independently attributed target flows. We train and evaluate three representative methods~\cite{etbert2022,xgboost2016,yatc2023} on this SNI-filtered subset, emulating evaluation on public benchmarks. However, expanding evaluation to all independently attributed target flows reduces balanced accuracy (BA) by 12.99--42.87 percentage points. These results show that overstrict filtering undermines benchmark \textit{representativeness}, as high scores on the filtered subset may not reflect model performance on all target flows.

\begin{figure}[tb]
\centering
\includegraphics[width=0.8\linewidth]{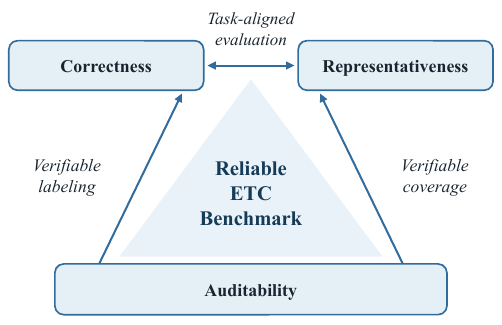}
\caption{CRA requirements for ETC benchmarks.}
\label{fig:cra}
\end{figure}

Building on prior research on benchmark quality and reliability\cite{iso5259,paullada2021,wangstrong1996,pipino2002}, our findings motivate three fundamental requirements for reliable ETC benchmarks: \textbf{Correctness, Representativeness, and Auditability (CRA)}. Correctness requires that each flow be labeled according to the target it actually belongs to\cite{iso25012,iso25024}. Representativeness requires the evaluation set to cover the declared set of target flows\cite{ransohoff1978,buolamwini2018}. We treat auditability as the basis for independently verifying both requirements (Fig.~\ref{fig:cra}). A dataset release should preserve the evidence behind each label and the evidence and rules supporting sample selection\cite{klie2024}. These checks help researchers judge whether higher benchmark scores reflect better performance on the intended classification task\cite{taori2020natural}.

To our knowledge, this is the first systematic audit of the complete lifecycle of ETC datasets. Our main contributions are as follows:
\begin{enumerate}
\item \textbf{Lifecycle systematization.} We develop a taxonomy covering collection, labeling, and downstream use, and systematically audit 13 public datasets. This extends dataset auditing beyond post-release processing and evaluation to upstream collection and label construction.
\item \textbf{Identification of construction defects.} We identify coarse inheritance and overstrict filtering as two key problems. Experiments show that these practices can conceal losses in target recognition and mislead judgments about ETC performance and method improvements.
\item \textbf{An independently attributed reference corpus.} We construct \vt, a corpus of traffic from website visits. Associations among visits, requests, connections, and network flows provide verifiable per-flow labels. 
\item \textbf{Dataset construction principles.} We derive practical recommendations for dataset construction, release, and evaluation from the requirements of correctness, representativeness, and auditability.
\end{enumerate}

\section{A Taxonomy of ETC Datasets}
\label{sec:taxonomy}

\begin{figure}[t]
\centering
\includegraphics[width=0.9\linewidth]{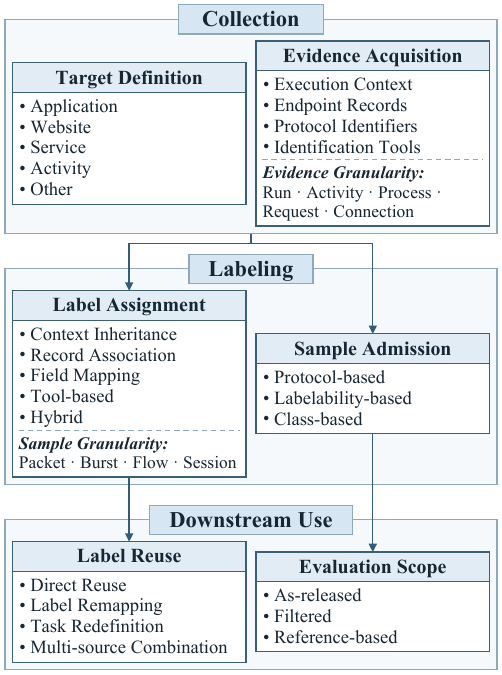}
\caption{ETC dataset lifecycle taxonomy.}
\label{fig:taxonomy}
\end{figure}

We divide the ETC dataset lifecycle into three stages---\emph{collection}, \emph{labeling}, and \emph{downstream use}---and organize the key decisions at each stage (Fig.~\ref{fig:taxonomy}).

Throughout the lifecycle\cite{iso5259part1}, we examine how datasets assign flow labels, select traffic for evaluation, and preserve evidence to verify these decisions. These three aspects correspond to correctness, representativeness, and auditability, respectively.

\subsection{Collection}
\label{sec:collection}
Collection defines the object of study and obtains raw traffic together with labeling evidence. We divide these decisions into target definition and evidence acquisition.

\textbf{Target definition.} ETC tasks can be grouped by the object they seek to identify:

\textbf{Application identification} identifies the application generating the communication~\cite{mirage2019,nudt2024}. The label denotes application identity and need not describe a specific user action.

\textbf{Website-visit identification} identifies the website or page a user visits~\cite{sok2025}. One visit may generate multiple connections and retrieve resources from different servers.

\textbf{Service identification} identifies the network service associated with each connection, often using domain information such as SNI~\cite{cesnettls2023}. The identified service may differ from the website the user is visiting.

\textbf{Activity identification} identifies user activities such as chatting, transferring files, or playing videos. One application can support multiple activities, and one activity may involve multiple applications or connections~\cite{iscxvpn2016}.

\textbf{Other classification tasks} include identifying VPN or Tor encapsulation~\cite{iscxvpn2016,iscxtor2017}, distinguishing benign from malicious traffic~\cite{doh2020}, and classifying attacks. Label semantics follow the objective of each task.

These tasks can use the same raw traffic~\cite{etbert2022}, but they ask different questions: which application generated a flow, which service it connects to, or which website visit it belongs to. Each question therefore requires evidence linking the flow to the corresponding target.

\textbf{Evidence acquisition.} We distinguish four sources of evidence used to establish labels:

\textbf{Execution context} records the application, activity, or experimental conditions of a collection run. Such evidence usually describes an entire run or activity~\cite{iscxvpn2016,iscxtor2017}.

\textbf{Endpoint records} capture processes, requests, connections, or other execution information from a host or application. They can associate applications with network communication at different granularities~\cite{mirage2019,nudt2024,netlog2020}.

\textbf{Protocol identifiers} use information carried by network protocols, such as SNI in a TLS handshake, to characterize the remote endpoint~\cite{rfc6066,cesnettls2023}.

\textbf{Identification tools} use the outputs of deep packet inspection (DPI) or existing classification systems as labeling evidence~\cite{appclassnet2022}.

 We refer to the level at which evidence applies as its \textbf{evidence granularity}\cite{carata2014}. Common granularities include runs, activities, processes, requests, and connections.
 
\subsection{Labeling}
\label{sec:labeling}
Labeling converts collection evidence into labels for network samples and selects samples for the final dataset. The main decisions concern label assignment and sample selection.

\textbf{Label assignment.} Evidence acquisition determines what information is available for labeling. Label assignment determines how that information is used to label individual flows.
We distinguish the following ways of transferring labels from source evidence to network samples:

\textbf{Context inheritance} assigns the label of a collection run or activity to its network samples. The sample's collection context determines its class~\cite{iscxtor2017}.

\textbf{Record association} establishes labels through correspondences between endpoint and network records. These associations can use connection identifiers, request records, or other verifiable information~\cite{mirage2019,a3c2022,gringoli2009gt}.

\textbf{Field mapping} derives classes from protocol fields and predefined mappings, for example by mapping server names to service labels~\cite{cesnettls2023}.

\textbf{Tool-based labeling} adopts the classes produced by an existing identification tool~\cite{appclassnet2022}.

\textbf{Hybrid labeling} combines these approaches using predefined priorities or fallback rules~\cite{mirage2019}.

Label assignment also involves \textbf{sample granularity}. Common units include packets, bursts, flows, and sessions, whose definitions have been systematized in prior work~\cite{sok2025}.

\textbf{Sample selection.} Dataset builders usually select part of the collected traffic for labeling and release. We distinguish three types of selection by the criterion used to retain or discard samples:

\textbf{Protocol-based selection} selects samples by protocol type, encryption form, or connection state to define the dataset's communication scope~\cite{etbert2022,cesnettls2023}.

\textbf{Labelability-based selection} retains samples according to whether a label can be obtained\cite{lakkaraju2017}. Examples include requiring a parsable protocol field, an identifier that maps to a target class~\cite{cesnettls2023}, or a corresponding execution record.

\textbf{Class-based selection} filters samples by the class set, class frequency, or sample quotas to adjust class and sample distributions~\cite{etbert2022,appclassnet2022}.

These criteria can be applied sequentially to the same collected traffic. The final release contains the retained samples and their labels. Labeling evidence may also be preserved, either implicitly in the samples themselves or explicitly in accompanying records.

\subsection{Downstream Use}
\label{sec:downstream}
Downstream research builds classification tasks from released datasets and compares methods through model training and testing~\cite{yatc2023}. We group its main choices into label reuse and evaluation scope.

\textbf{Label reuse.} Original labels enter downstream tasks in several ways:

\textbf{Direct reuse} adopts the released labels and class definitions~\cite{guarino2023few}.

\textbf{Label remapping} reorganizes existing classes without changing the classification target~\cite{cesnettls2023}.

\textbf{Task redefinition} reuses existing traffic to identify a different target~\cite{etbert2022,netbench2024}.

\textbf{Multi-source combination} combines datasets into a unified task and organizes labels from different sources~\cite{darknet2020,flowpic2021}.

These practices can change label organization or semantics, but they do not add collection evidence for the original labels.

\textbf{Evaluation scope.} We distinguish evaluation settings by which samples actually contribute to the reported score:

\textbf{As-released evaluation} trains, tests, and compares methods within the scope of the original release.

\textbf{Filtered evaluation} further selects released samples by protocol, class, or labeling conditions and evaluates on the resulting subset~\cite{cesnettls2023,mirage2019,wang2024fssl}.

\textbf{Reference-based evaluation} constructs an evaluation set from target samples with independent attribution evidence to examine performance on the original release or a filtered subset.

Apart from sample scope, training and testing can be split by individual sample, collection run, or time~\cite{sok2025,cesnettls2023}. Splitting controls relationships between training and test data\cite{sogaard2021splits,arp2022dos}; evaluation scope determines which target samples are scored\cite{koh2021wilds,gardner2023tableshift}.

The key lifecycle decisions thus determine \textbf{how labels are produced, which samples are retained, and how those labels and samples enter downstream tasks.} We use this taxonomy to organize the construction and reuse of 13 public ETC datasets and examine their correctness, representativeness, and auditability.
\section{Systematization of ETC Datasets}
\label{sec:systematization}

\subsection{Dataset Discovery and Selection}
Our study covers benchmark datasets used in ETC research and their derived tasks. We compiled candidates from existing surveys, SoK papers, and audit inventories. We identified downstream studies through dataset names, aliases, and forward citations to source papers. The first automated search was conducted on August 2, 2026, covering literature published from January 1, 2020, through July 31, 2026. We used OpenAlex and Semantic Scholar to locate sources and supplement metadata, and the Semantic Scholar API for forward-citation and keyword searches. Candidate papers were deduplicated by DOI and normalized title, while preserving relationships between original datasets and derived versions\cite{longpre2024,halevy2016goods,huang2017orpheus}.

To focus on academically influential ETC datasets, we required the canonical source paper to have more than 30 Google Scholar citations. For eligible entries, we collected source papers, official documentation, code, released data, and downstream papers to verify construction and use. We included only datasets with publicly accessible data or relevant release materials. Entries that were no longer publicly available and lacked verifiable data were excluded from the final audit set.

\begin{table*}[t]
\caption{Lifecycle classification matrix of ETC datasets.}
\label{tab:lifecycle}
\centering
\setlength{\lifecyclewidth}{\dimexpr\textwidth-14\tabcolsep\relax}
\begin{tabular}{@{}
L{0.150\lifecyclewidth}
L{0.105\lifecyclewidth}
L{0.125\lifecyclewidth}
L{0.135\lifecyclewidth}
L{0.105\lifecyclewidth}
L{0.120\lifecyclewidth}
L{0.120\lifecyclewidth}
L{0.140\lifecyclewidth}@{}}
\hline
\textbf{Dataset} &
\textbf{Target} &
\textbf{Label evidence} &
\textbf{Assignment} &
\textbf{Selection} &
\textbf{Label reuse} &
\textbf{Evaluation} &
\textbf{Verification} \\
\hline

ISCXVPN2016~\cite{iscxvpn2016} &
App., activity, VPN &
Run context &
Context inheritance &
NE &
Derived app./service tasks &
Released; protocol-filtered &
PCAPs; collection docs. \\
\hline

ISCXTor2016~\cite{iscxtor2017} &
Activity, Tor &
Run context &
Context inheritance &
NE &
CIC-Darknet reuse~\cite{darknet2020} &
Derived set &
PCAPs; collection docs. \\
\hline

USTC-TFC2016~\cite{ustctfc2016} &
App., malware &
Upstream class records &
Source-label inheritance &
Source datasets combined &
Direct reuse &
Released; protocol-filtered &
PCAPs; upstream docs. \\
\hline

MIRAGE-2019~\cite{mirage2019} &
Application &
System-call; connection logs &
Exact assoc.; majority fallback &
NE &
Direct reuse &
Released; \texttt{exact} subset &
Flow labels; assignment sources \\
\hline

CIC-Darknet2020~\cite{darknet2020} &
Activity, encapsulation type &
Upstream labels &
Inheritance; class reorganization &
Source combination &
Derived reorganization &
Combined set &
Source links; flow features \\
\hline

CIRA-CIC-DoHBrw~\cite{doh2020} &
DoH use, maliciousness &
Run scenarios &
Context inheritance &
NE &
Hierarchical tasks~\cite{mitsuhashi2023doh} &
Released &
PCAPs; run docs. \\
\hline

CSTNET-TLS1.3~\cite{etbert2022} &
Remote domain &
Connection SNI &
Field mapping &
TLS 1.3 target set &
Direct reuse &
Released &
Per-flow PCAPs; labels \\
\hline

ET-BERT-derived tasks~\cite{etbert2022} &
App., service &
Upstream labels &
Label inheritance; reorganization &
Per-class quotas &
Derived tasks &
Derived subsets &
Code; derived data \\
\hline

CESNET-TLS22~\cite{cesnettls2023} &
Network service &
Connection SNI &
Service mapping &
TLS; service classes &
Service identification &
Released &
Service maps; stats. \\
\hline

CESNET-QUIC22~\cite{cesnetquic2023} &
Service, background class &
QUIC SNI &
Service mapping &
QUIC, SNI; sampling &
Service/provider tasks &
Sampled set &
Service maps; sampling stats. \\
\hline

AppClassNet~\cite{appclassnet2022} &
Application ID &
Commercial DPI outputs &
Tool inference &
Frequent classes &
Direct reuse &
Released &
Anonymized features; construction docs. \\
\hline

CipherSpectrum~\cite{sok2025} &
Visit target, remote domain &
Visit records; SNI &
Visit association; field mapping &
TLS 1.3; cipher-suite quotas &
Two label views (this study) &
Corresponding target sets &
PCAPs; collection-target records \\
\hline

NUDT MobileTraffic~\cite{nudt2024} &
Application &
App.--connection logs &
Record association &
TCP; DNS &
NE &
Associated subset (this study) &
PCAPs; app.--connection logs \\
\hline

\multicolumn{8}{@{}l}{\textit{NE: not established;
App.: application; assoc.: association;
docs.: documentation; stats.: statistics.}} \\
\hline
\end{tabular}
\end{table*}

For each dataset, we recorded the classification target, labeling evidence, label assignment, and sample selection from the source materials. We traced label reuse and evaluation scope through downstream papers~\cite{guarino2023few,luxemburk2023quic}. We distinguished actual training or testing from comparisons that reused published scores and from background citations. Where public data and processing records were available, we also examined per-flow labels against the release materials. Where only textual descriptions were available, we recorded only the construction information those descriptions established.

\subsection{Lifecycle Classification of the Datasets}
We organize the construction and reuse of 13 ETC datasets using the taxonomy in Section~\ref{sec:taxonomy}. Table~\ref{tab:lifecycle} presents the lifecycle classification matrix. It summarizes classification targets, labeling evidence, label assignment, sample selection, label reuse, and evaluation scope. Labeling evidence identifies the basis used to assign labels; it need not establish attribution for each individual flow. The table also records the label-verification materials available in the public releases.

Table~\ref{tab:lifecycle} shows that ETC datasets differ not only in their labeling methods, but also in \textbf{which classification targets the original evidence directly supports and how that evidence is converted into per-flow labels}. Application labels can originate from collection activities, process--connection records, or existing identification tools. These sources differ in how directly they support per-flow attribution, even when the resulting labels have the same form~\cite{dusi2011}.

\textbf{The distinction between collection context and per-flow attribution is already evident in early ETC datasets.} The original ISCXVPN2016 study organized collection around activities such as browsing, chat, video streaming, and VoIP, generating both ordinary encrypted traffic and VPN traffic. The paper notes that Hangouts voice-call captures also contained browsing flows~\cite{iscxvpn2016}. ISCXTor2016 used an explicit context-based assignment procedure. After confirming that most flows in an ordinary capture came from the target application, the authors assigned that application's label to every flow in the corresponding Tor capture. They also acknowledged that flows from other sources would inherit the target label, introducing label noise~\cite{iscxtor2017}. These procedures use collection conditions to provide supervision. However, these datasets do not establish whether each connection was actually generated by the application associated with its collection activity.

\textbf{Endpoint records provide a more direct route to per-flow attribution.} MIRAGE uses process and connection information from system-call logs to associate network five-tuples with Android applications. Matched flows receive exact labels. Unmatched flows receive the most frequent application label in the capture, and the release distinguishes these two assignment sources~\cite{mirage2019}. NUDT MobileTraffic uses Netlog to record application labels, timestamps, and connection five-tuples. Linking these records to network flows allows application--flow relationships to be reconstructed~\cite{nudt2024}. These construction methods illustrate the role of endpoint evidence: collectors can directly record the relationship between an application and its communication, while distinguishing exact associations from inferred assignments.

\textbf{Evidence from network identifiers must be interpreted in relation to the classification target.} CSTNET-TLS1.3 uses SNI to label connections~\cite{etbert2022}. CESNET-TLS22 maps SNI values to service names to construct a network-service identification task. Its source paper explicitly discusses the incomplete coverage of service-domain mappings~\cite{cesnettls2023}. SNI supplies the remote name declared for a connection; a service mapping then assigns that name to a service category. SNI can identify the remote domain associated with a connection, but does not establish which website visit generated it. Website-visit labeling therefore requires additional evidence linking connections to their corresponding visits.

These differences show that \textbf{dataset construction requires joint scrutiny of label meaning, evidence granularity, and sample-selection criteria}. Run context, endpoint records, protocol fields, and identification tools can supply labels for different tasks. The meaning of their evidence must correspond to the declared classification target. Dataset builders may also filter flows by protocol type or require identifiers such as SNI. Together, label assignment and sample selection define the task that downstream models learn and are evaluated on. They also provide the basis for our analysis of coarse inheritance and overstrict filtering.

\section{Key Problems in ETC Dataset Construction}
\label{sec:problems}

We examine two problems in dataset construction. \textbf{Coarse inheritance} occurs when context inheritance (Section~\ref{sec:labeling}) transfers a collection-run or activity label to individual flows without corresponding per-flow attribution evidence. \textbf{Overstrict filtering} excludes flows belonging to the intended classification task because they fail a selection rule. These problems concern label correctness and coverage of target flows, respectively. We further examine which verification materials survive release and formulate the research questions addressed in Section~\ref{sec:evaluation}.

\subsection{Problem 1: Coarse Inheritance}
\label{sec:coarse-problem}
Some ETC datasets collect traffic by executing a particular application or user activity, then assign the recorded class to every flow~\cite{sugar2025,etbert2022,iscxvpn2016}. This converts a run- or activity-level label into per-flow application labels. Its correctness depends on an important assumption: every connection in the capture belongs to the recorded target application\cite{dietterich1997,zhou2018weak}.

A packet capture collected during an application run may also contain unrelated background traffic from system services or other applications~\cite{zaki2021grano}. Collection records can establish which activity was performed without establishing which application generated each connection~\cite{mirage2019,nudt2024}. When run labels are used directly for flow classification, these attribution differences may be collapsed into a single class. ISCXVPN and its derived tasks illustrate this inheritance from collection context to per-flow labels~\cite{etbert2022,iscxvpn2016,distiller2021}.

  The problem extends to downstream method evaluation. Downstream studies do not generally treat these datasets as datasets with noisy labels~\cite{lowquality2024}. Training asks models to learn the inherited labels, and testing judges predictions against the same labels~\cite{sugar2025}. For connections whose actual source differs from the collection label, a model that correctly identifies the source may be marked wrong\cite{vasudevan2022}. A model that better reproduces the collection labels may score higher\cite{geva2019,obermeyer2019}. Even a strict train/test separation therefore does not guarantee that evaluation measures the intended application-identification task.

If two flows have different application labels, the model is required to distinguish them. If their input features are identical, however, it cannot classify both correctly~\cite{jerabek2025}. Identical model inputs with different application labels raise two questions: whether the labels are supported by the collection evidence, and whether the input contains enough information to distinguish the flows. We examine these questions separately and compare input conflicts before and after protocol filtering~\cite{sok2025,sugar2025}.

\begin{researchquestions}{Coarse Inheritance --- Research Questions}
\noindent\textbf{RQ1:} How does assigning the same application label to all flows in a capture affect model training and classification performance?
\par\smallskip
\noindent\textbf{RQ2:} What happens when flows with different application labels produce identical model inputs?
\end{researchquestions}

\subsection{Problem 2: Overstrict Filtering}
\label{sec:filtering-problem}
Another problem arises when network-side identifiers determine labels. Fields such as SNI provide the server name of a connection's remote endpoint~\cite{rfc6066}, making them widely used in service- and application-classification datasets~\cite{sok2025,etbert2022,cesnettls2023,cesnetquic2023}. However, \textbf{connections identifiable through SNI need not cover all communications generated by an application or website.} There are two main reasons.

\textbf{First, even visible SNI may not match the visit target.} A website visit typically creates multiple connections to retrieve pages, images, scripts, and other resources from different servers~\cite{what2016,englehardt2016tracking}. Each connection's SNI names its own server, rather than the common visit target it supports. Retaining only connections whose SNI matches the target website can therefore exclude flows to other resource servers. Those flows still belong to the visit and should be included in website-visit identification, even though their SNI differs from the target domain.

  \textbf{Second, some encrypted communications do not expose target SNI suitable for matching.} SNI is not mandatory in every TLS connection, and its availability also depends on the protocol and capture completeness. Encrypted Client Hello (ECH) can encrypt the actual server name in a TLS connection; an externally visible SNI may identify only a public service entry point~\cite{rfc9849}. Other encrypted communications, such as E2EE~\cite{mls}, do not necessarily use SNI from an ordinary TLS connection. Labeling rules that depend on visible network identifiers cannot cover all these forms of communication.

Both cases can cause \textbf{connections belonging to the classification target to be excluded because they fail the labeling rule}. If labelability also determines sample selection, the final dataset may contain only target flows that are easier to label. 

This selection also affects evaluation. When the same rule determines both training and test samples, a model only needs to perform well on retained connections. Recognition failures on other target connections do not affect its reported score. A filtered benchmark may therefore overestimate performance on the intended classification task and distort comparisons between methods\cite{tesseract2019,krichene2020sampled}.

\begin{researchquestions}{Overstrict Filtering --- Research Questions}
\noindent\textbf{RQ3:} When network identifiers such as SNI determine labels and sample selection, which target flows are excluded? How does this selection affect coverage of target flows?
\par\smallskip
\noindent\textbf{RQ4:} When the same selection rule is used for training and evaluation, how do benchmark scores differ from performance on the intended classification task? How does this difference affect assessments of models trained on different training sets?
\end{researchquestions}

\subsection{Auditability: Can Construction Problems Be Independently Verified?}
\label{sec:auditability-problem}
Both problems concern the relationship between collection evidence and the final dataset. Examining coarse inheritance requires associations between original activities and individual connections. Examining overstrict filtering requires evidence identifying target flows and showing which were retained or discarded. Public datasets, however, mainly release network samples and final labels; they do not necessarily preserve sufficient attribution evidence or selection histories~\cite{sugar2025,mirage2019}.

Without this evidence and documentation, downstream researchers can still train models and reproduce scores, but may be unable to establish why a flow received a particular label or which target connections were excluded during labeling. Resplitting, class mapping, and repeated reuse can perpetuate this difficulty~\cite{sugar2025,etbert2022}.

Auditability thus requires more than public availability\cite{yang2024datasetcards}. Released materials must support reconstruction of how labels and sample scope were formed~\cite{datasheets2021,datacards2022,hutchinson2021,jo2020archives}.

\begin{researchquestions}{Auditability --- Research Question}
\noindent\textbf{RQ5:} To what extent do the public materials of existing ETC datasets support independent verification of per-flow labels and sample-selection processes?
\end{researchquestions}

Section~\ref{sec:evaluation} addresses these questions empirically. Together, they examine how dataset construction affects the evaluation of ETC methods.

\section{Experiments and Guidelines}
\label{sec:evaluation}

Our experiments show that dataset construction can make benchmark scores misleading measures of performance on the intended classification task. We first examine how coarse inheritance affects classification and how identical network inputs with different labels constrain model performance (RQ1 and RQ2). We then use \vt{} to investigate how network-identifier filtering affects coverage of target flows and method evaluation (RQ3 and RQ4). Finally, we examine whether released evidence allows independent verification of flow labels and sample selection (RQ5).

\subsection{VisitTrace: Collection and Attribution}
\label{sec:setup}

\begin{figure*}[t]
\centering
\includegraphics[width=0.95\linewidth]{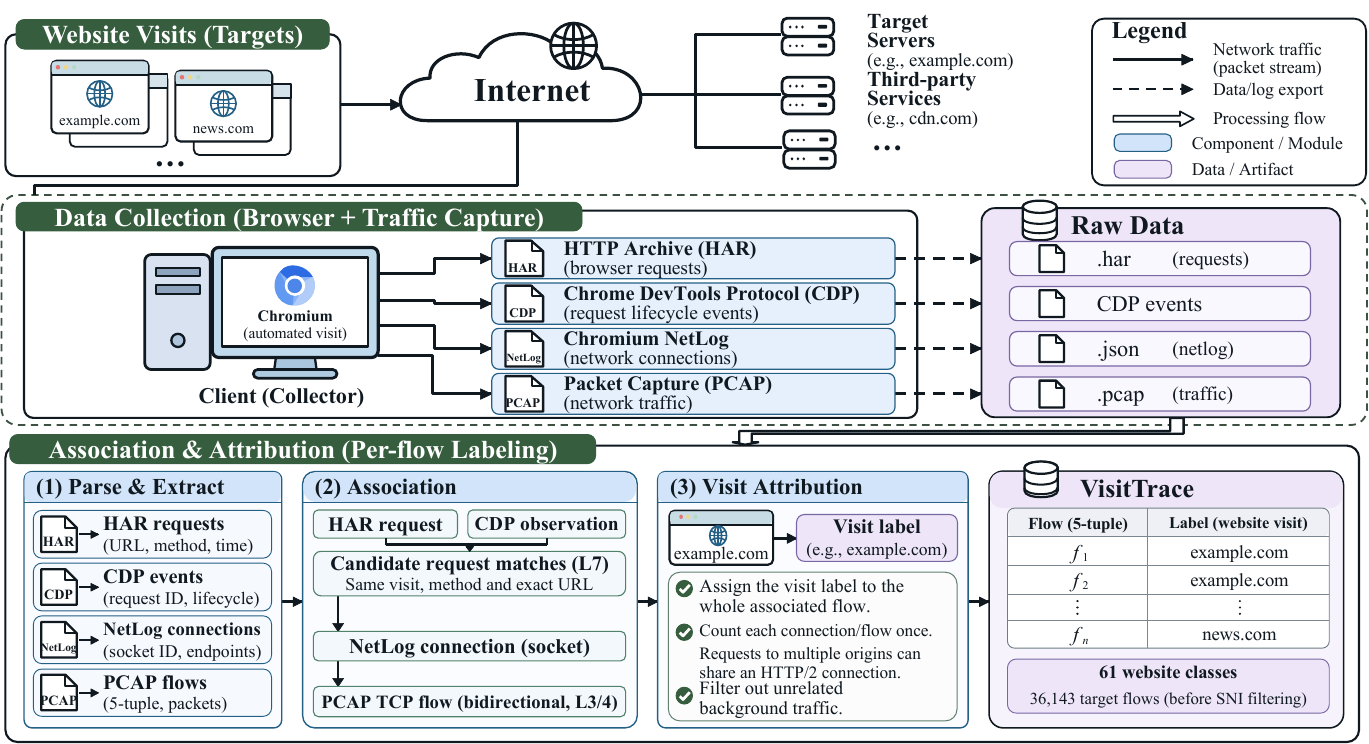}
\caption{The VisitTrace collection workflow.}
\label{fig:workflow}
\end{figure*}

We construct \vt{}, a corpus of network traffic captured
during real website visits. For each visit, we collect
HTTP Archive (HAR) files, Chrome DevTools Protocol
(CDP) events, Chromium NetLog records, and packet
captures (PCAP). HAR provides HTTP request and response
information, while CDP records request lifecycle events
and identifiers used to associate browser requests with
network connections. NetLog provides socket and connection
information, and PCAP contains the captured network traffic.
The website being visited defines the classification target.

To establish per-flow labels, we associate HAR requests
with CDP observations and use the available request,
response, and socket evidence to link them to NetLog
connections. We then match these connections to
bidirectional TCP flows in PCAP. Each flow with a
verified association is assigned the label of its
corresponding website visit. This process also distinguishes
target flows from unrelated background traffic. A flow
is not assigned the visit label merely because it was
observed during that visit.

A website visit may involve requests to multiple remote servers, including third-party services. Moreover, HTTP/2 allows multiple requests to share a connection~\cite{http2}. We therefore associate flows with their originating visits rather than labeling them according to the domains of their remote servers. When multiple requests share a connection, the corresponding network flow is counted only once. The prediction unit remains the network flow, while its label identifies the website visit to which it belongs.

This attribution does not rely on Server Name Indication (SNI). Connections can retain their visit labels even when their SNI differs from the visited website's domain or cannot be recovered. We can therefore apply SNI-based filtering separately and compare the selected flows against an independently established reference. All VisitTrace analyses use the same 61 website classes, comprising 36,143 independently attributed target flows.

\subsection{Coarse Inheritance (RQ1 and RQ2)}
\label{sec:coarse-evaluation}
\textbf{Coarse inheritance can affect both model training and performance evaluation.} We use MIRAGE because its system-call and connection records establish which application generated each matched flow. This lets us compare per-flow labels with controlled majority assignment on the same flows and against the same reference (RQ1). We then examine how identical network inputs with different application labels limit classification performance, and how protocol filtering changes these limits (RQ2).

Accuracy weights test flows equally, whereas balanced accuracy (BA) weights class recall equally. We use both to distinguish aggregate performance from recognition across classes.

\subsubsection{Effects on Model Training and Evaluation}
\label{sec:mirage}

\textbf{Coarse inheritance can make a model appear more accurate while weakening its ability to recognize the original application classes.} We examine this effect using MIRAGE, which provides independently attributed application labels for individual flows.

We use MIRAGE's matched per-flow labels as the reference~\cite{mirage2019}. To examine the effects of coarse inheritance, we apply a controlled majority assignment: each flow's label is replaced with the most common application among the other matched flows in the same capture. This procedure differs from ISCXVPN's direct inheritance of collection-activity labels. The experiment includes 84,990 flows from 1,639 captures, covering 47 applications. We train logistic regression (LR) and a lightweight one-dimensional CNN using both labeling schemes, with identical network inputs and three-fold splits grouped by capture. All models are evaluated against the original 47-class per-flow labels, and CNN experiments are repeated with three seeds.

A capture may contain flows from multiple applications. Under majority assignment, flows from less frequent applications receive the label of the most common application in that capture. Although only 4.54\% of flow labels change, 23 of the original 47 application classes disappear from the training labels. These applications account for just 0.58\% of all flows. Their connections remain in the dataset, but their original application labels no longer appear in training.

Both models achieve higher accuracy but lower balanced accuracy (BA) when trained with majority labels (Table~\ref{tab:mirage}). Accuracy gives more weight to frequent applications because they contribute more test flows, whereas BA weights every application class equally. Consequently, majority assignment appears preferable under accuracy, even though it produces lower BA on the same original classification task.

A per-fold decomposition shows that the BA decrease is
driven by classes that lose all training labels, rather
than a general decline across classes. These classes
contribute 11.87 and 10.99 percentage points to the BA
decrease for LR and CNN, respectively, while improved
recall among classes retaining training labels offsets
1.41 and 4.88 percentage points. Thus, majority
assignment can improve recognition of retained classes
while eliminating supervision for many other classes.

\begin{table}[t]
\caption{Effects of majority assignment on classification
performance and BA decomposition.}
\label{tab:mirage}
\centering

\begin{tabular}{llrr}
\hline
\multicolumn{4}{l}{\textit{(a) Classification performance}} \\
\hline
Model & Training labels & Accuracy (\%) & BA (\%) \\
\hline
LR & Per-flow & 25.40 & 25.59 \\
LR & Majority & 30.43 & 15.13 \\
CNN & Per-flow & $21.73\pm0.37$ & $22.97\pm1.02$ \\
CNN & Majority & $34.46\pm0.41$ & $16.86\pm0.08$ \\
\hline
\end{tabular}

\smallskip

\begin{tabular}{lrrr}
\hline
\multicolumn{4}{l}{\textit{(b) Decomposition of BA decrease (pp)}} \\
\hline
Model &
\shortstack{Lost-label\\classes} &
\shortstack{Retained-label\\classes} &
\shortstack{Net BA\\decrease} \\
\hline
LR & $+11.87$ & $-1.41$ & $10.46$ \\
CNN & $+10.99\pm0.82$ & $-4.88\pm0.18$ &
$6.11\pm0.95$ \\
\hline
\end{tabular}
\end{table}

\takeaway{1}{A small overall relabeling rate can conceal the complete loss of supervision for many targets. In MIRAGE, majority assignment raises accuracy while lowering BA on the original task. The proportion of changed labels alone cannot reflect the impact of coarse inheritance, as entire application classes may disappear from the training labels.}

\guideline{1}{Dataset builders should preserve the attribution evidence supporting each sample's label and clearly identify labels assigned from collection-level evidence or majority assignment. Downstream researchers should examine these labeling methods before reusing the dataset, and distinguish inferred labels from independently verified ones when interpreting model performance. }

\subsubsection{Input Conflicts and Classification Requirements}
\label{sec:conflicts}
\textbf{Coarse inheritance can assign different application labels to flows with identical network inputs.} Since a model cannot distinguish these flows using the same input, it cannot classify all of them correctly. This limitation applies regardless of which classifier is used.

We adopt Wickramasinghe et al.'s CTD occlusion strategy~\cite{sok2025},
which randomizes identifying information (including MAC/IP addresses,
ports, and SNI) as well as context- and time-dependent header fields.
We then construct two packet-window representations using the first
five packets (H5) and the first 20 packets (H20) of each flow. For short flows, all available packets are used and the remaining positions are padded. Flows with identical features form a group. A group containing multiple labels is a \emph{conflict group}; the \emph{conflict-flow fraction} is the number of flows in such groups divided by the size of the current set.

Choosing the class that maximizes BA within each identical-feature group yields the empirical BA ceiling for the given input. Let $n_{g,c}$ be the number of flows with label $c$ in group $g$, $n_c$ the total number of flows in class $c$, and $K$ the number of classes. Then
\begin{equation}
\operatorname{BA}_{\max}=\frac{1}{K}\sum_g\max_c\frac{n_{g,c}}{n_c}.
\label{eq:ba-ceiling}
\end{equation}

ISCXVPN shows that these conflicts can characterize most of a released task, rather than a few exceptional samples (Table~\ref{tab:protocol}). Its per-flow application task assigns capture-level classes to individual flows~\cite{etbert2022,iscxvpn2016}. Common network behavior recurring across captures consequently receives different application labels despite identical H20 inputs. Before protocol filtering, the empirical BA ceiling is 62.37\%, bounding performance even for a model that fully exploits the specified representation. The empirical BA ceiling diagnoses incompatible supervision under that representation; attribution evidence is needed to establish what the conflicting labels describe.

\takeaway{2}{Assigning coarse labels to all flows in a capture can give different labels to flows with identical network inputs. No classifier can correctly distinguish such flows using the same inputs, regardless of its capacity. Reliable classification therefore requires examining how labels are assigned, rather than relying solely on more powerful models.}

\subsubsection{Protocol Filtering Changes the Classification Task}
\label{sec:protocol-cleaning}

Protocol filtering can raise the empirical BA ceiling by removing traffic from selected protocols, even when all classes are retained. This changes the classification task itself, rather than improving recognition of the original traffic.

We apply the protocol-filtering rules of Zhao et al.~\cite{sugar2025}, preserve the original labels, re-extract H5/H20 features, and recompute input conflicts and empirical BA ceilings. Table~\ref{tab:protocol} reports sample scope and classification requirements before and after protocol filtering, with H20 conflicts expressed as the conflict-flow fraction.

ISCXVPN provides the clearest example of an easier retained task: protocol filtering removes 81.82\% of flows while preserving every application class, raising the H20 empirical BA ceiling from 62.37\% to 84.38\%. The two largest original conflict groups mainly contain two-packet Link-Local Multicast Name Resolution (LLMNR) exchanges shared across multiple application labels. Filtering removes much of this common behavior. The increase in the empirical BA ceiling reflects a narrower evaluation set.

Increasing the packet window cannot resolve conflicts in flows that contain only one packet. After protocol filtering, the largest remaining conflict group in ISCXVPN consists of single-packet UDP flows with different application labels. Across the retained flows, extending the packet window from H5 to H20 raises the empirical BA ceiling by only 0.65 percentage points.

\textbf{Protocol filtering must be guided by the classification task.} LLMNR, NetBIOS Name Service (NBNS), and ICMP traffic may provide little useful information for application identification, so excluding them can be reasonable for that task. However, LLMNR and NBNS traffic can be involved in name-resolution poisoning, while ICMP traffic can carry malicious commands~\cite{mitre-t1557,mitre-t1095}. Removing these flows from datasets intended to detect such attacks would exclude relevant traffic from evaluation. Thus, whether a protocol should be filtered depends on the intended classification task, rather than a universal exclusion rule.

\takeaway{3}{Protocol filtering can increase the empirical BA ceiling by removing flows from selected protocols. This changes the classification task itself. A higher empirical BA ceiling does not mean the original task has become easier to solve.}

\guideline{2}{Protocol filtering is a temporary workaround for labeling problems in existing datasets. Future datasets should establish and preserve reliable per-flow attribution evidence during collection. When filtering existing datasets, researchers should document the rules and preserve evidence of sample selection, including which flows were excluded, allowing others to verify how filtering changes the classification task.}

\begin{table*}[t]
\caption{Input conflicts and empirical BA ceilings before/after protocol filtering (Both: unchanged).}
\label{tab:protocol}
\centering
\begin{tabular}{L{0.26\textwidth}lrrrrr}
\toprule
Dataset / label object & Filtering & Flows & Classes &
\shortstack{H5 empirical\\BA ceiling} &
\shortstack{H20 empirical\\BA ceiling} &
H20 conflicts \\
\midrule

ISCXVPN / application
 & Before & 310,141 & 16 & 61.87\% & 62.37\% & 94.047\% \\
 & After  & 56,388  & 16 & 83.73\% & 84.38\% & 69.985\% \\

\midrule
ISCXVPN / activity
 & Before & 310,141 & 6 & 62.97\% & 62.99\% & 91.969\% \\
 & After  & 56,388  & 6 & 83.21\% & 83.31\% & 58.755\% \\

\midrule
USTC / 20 released labels
 & Before & 595,777 & 20 & 87.74\% & 88.10\% & 41.052\% \\
 & After  & 595,635 & 20 & 87.74\% & 88.10\% & 41.055\% \\

\midrule
CSTNET / 120 released labels
 & Both & 46,369 & 120 & 99.29\% & 99.99\% & 0.013\% \\

\midrule
CipherSpectrum / visit target
 & Both & 123,000 & 85 & 50.75\% & 97.34\% & 7.166\% \\

\midrule
CipherSpectrum / remote domain
 & Both & 123,000 & 46 & 65.61\% & 99.82\% & 0.518\% \\

\midrule
NUDT / endpoint-attributed application
 & Both & 6,566,838 & 350 & 42.01\% & 90.46\% & 18.817\% \\

\midrule
\vt{} / website visit
 & Both & 36,143 & 61 & 93.52\% & 99.44\% & 1.65\% \\

\bottomrule
\end{tabular}
\end{table*}

\subsubsection{Independent Attribution and Input Separability}
\label{sec:independent-attribution}
\textbf{Independently attributed flows can still have identical network inputs.} In NUDT, different applications can generate identical TCP control flows without application payload. In \vt{}, visits to different websites can contact the same third-party service, producing flows with identical features in their first 20 packets. Independent attribution evidence confirms that these flows belong to different targets, even though the available network inputs cannot distinguish them.

In NUDT MobileTraffic, application--connection logs establish the application generating each connection. In one H20 conflict group, flows attributed to 321 applications each consist of six TCP control packets without layer-4 (L4) payload and share identical H20 features. These are application-specific connections exhibiting common connection-establishment and termination behavior. Their labels are supported by endpoint records even though their network inputs do not reveal the application identity.

In \vt, extending the packet window from H5 to H20 reduces the conflict-flow fraction from 23.36\% to 1.65\%, although some conflicts remain because different websites share the same services. Every remaining H20 conflict flow contains L4 payload and belongs to a complete flow longer than 20 packets. Inspection shows that visits to \url{docs.djangoproject.com}, \url{docs.python.org}, and \url{www.python.org} all generate requests to \url{analytics.python.org/api/event}. Visit--request--connection records confirm the respective visit targets. The first 20 packets have identical features, while the complete flows exhibit payload-length differences after packet 20. The visit labels thus describe genuine differences in origin that the truncated inputs conceal.

These cases point to different research responses. A longer packet window may distinguish the \vt{} flows because later packets contain additional information. It offers no additional packets for the short NUDT control exchanges. Zhao et al.'s protocol filtering leaves the main input-conflict findings for both datasets materially unchanged.

\takeaway{4}{Different applications or website visits can produce flows with identical network inputs, even when their labels are correct. These flows reveal limits in what a classifier can recognize from the available input. Removing them would change the classification task instead of solving the problem.}

\guideline{3}{Dataset builders should preserve evidence that establishes which target each flow belongs to. Downstream researchers should use this evidence to verify labels and distinguish labeling errors from cases where different targets produce identical network inputs. Correctly labeled flows should remain in evaluation, even when models cannot distinguish them.}

\subsection{Overstrict Filtering (RQ3 and RQ4)}
\label{sec:filtering-evaluation}
\textbf{Overstrict filtering can exclude flows belonging to the target.} When the same rule is applied to training and testing, models are trained and evaluated only on retained flows. A model can misclassify excluded target flows without penalty, causing benchmark scores to overestimate its performance on the intended classification task. We use \vt{} to measure how many target flows are excluded by SNI-based filtering and how this affects model evaluation.

\subsubsection{Per-Flow Reference and Experimental Design}
\label{sec:filter-design}
\textbf{Testing the coverage of a labeling rule requires target attribution established independently of that rule.} Website-visit identification assigns a connection to the visit that initiated it, whereas SNI describes the server name declared for that connection. CSTNET-TLS1.3 uses SNI to generate connection classes~\cite{etbert2022}. SNI identifies the remote domain associated with a connection, but does not establish which website visit generated it. CipherSpectrum illustrates this distinction: labeling the same flows by visit target and remote domain produces different input-conflict patterns (Table~\ref{tab:protocol}). Website-visit identification therefore requires additional evidence linking visits to their corresponding connections.

  Existing public ETC datasets\cite{etbert2022,sok2025} do not provide an independently attributed reference for target flows excluded during construction, making it impossible to quantify these omissions from the released data alone; VisitTrace provides the necessary pre-filtering reference. We apply SNI-based filtering to the independently attributed \vt{} flows (Section~\ref{sec:setup}), using the target-domain set frozen during collection. A flow is retained if its SNI equals a target domain or is its subdomain at a dot boundary.

We compare three training sets to distinguish the effects of sample selection and training size. FULL contains all reference training flows; RULE contains rule-matched training flows. RANDOM samples from FULL within each target and run, matching RULE's flow count in every group. We follow the experimental settings of Wickramasinghe et al.~\cite{sok2025}, including CTD occlusion. XGBoost~\cite{xgboost2016}, YaTC~\cite{yatc2023}, and ET-BERT~\cite{etbert2022} each run with three seeds on these sets, using the same predefined splits and complete reference validation set. Evaluation-scope comparisons hold models, predictions, and the 61 website classes fixed. Comparisons between training sets hold the target test set fixed. For paired intervals, we resample collection groups within website-target strata and average the differences across the three training seeds in each resample.

The experiment demonstrates the consequences of this
filtering mechanism without assuming that its measured
effects have the same magnitude in other ETC datasets.

\subsubsection{Coverage of Target Flows}
\label{sec:coverage}

\textbf{Overstrict filtering can exclude a substantial proportion of target flows, reducing benchmark representativeness.} Across \vt's common set of 61 website classes, the SNI rule retains only 13.58\% of independently attributed target flows. The excluded flows have the same kind of visit--request--connection evidence as retained flows. Their absence from the filtered corpus follows from the matching condition, rather than a lack of evidence that they belong to the target visits.

SNI-based filtering affects many websites, although its
coverage varies substantially across targets. Of the 61
websites, 38 lose more than half of their attributed flows.
The median per-website retention rate is 34.55\%, compared
with the flow-weighted rate of 13.58\%. Among the 31,235
excluded flows, 31,200 have visible but nonmatching SNI,
and 30,305 (97.02\%) carry only requests to registrable
domains different from that of the visited page.

The rule and the reference answer different questions. SNI matching asks whether a connection's declared server name falls within the specified domain set. Visit attribution asks which visit generated the connection. Resource connections outside that domain set can still belong to the visit. Excluding them equates membership in the website-visit identification task with satisfaction of the domain-matching rule. Consequently, overstrict filtering can narrow the scope of the intended classification task covered by those evaluation results.

Filtering by SNI or protocol type can exclude flows that belong to the classification target. For example, a website visit may contact third-party domains whose SNI does not match the visited website. Target flows may also use protocols outside the selected set, such as WireGuard or OpenVPN~\cite{wireguard,openvpn}. These flows can be excluded even though they belong to the intended classification task. 

\takeaway{5}{Overstrict filtering can exclude flows that genuinely belong to the classification target because they do not satisfy SNI or protocol-matching rules. These flows are then absent from evaluation, potentially causing benchmark scores to overestimate model performance on the intended classification task.}

\guideline{4}{Dataset builders should establish target-flow labels using independent attribution evidence, without relying on network-identifier matching to determine which flows belong to the target. They should preserve both attribution evidence and evidence of sample selection, including which target flows were excluded, so that filtering decisions can be independently verified. Downstream researchers should examine this evidence to assess whether the benchmark covers the target flows required by their classification tasks.}

\subsubsection{Evaluation Bias and Model Assessment}
\label{sec:evaluation-bias}
\textbf{Evaluating models only on SNI-matched flows can overestimate their performance on the intended classification task.} To examine this, we evaluate the same trained models on both the SNI-matched flows and all independently attributed target flows. The models remain unchanged; only the test flows differ. Table~\ref{tab:filter-results} reports the results for FULL, RULE, and RANDOM, averaged over three seeds. We define $\Delta$BA as the difference between filtered-test BA and full-reference BA, with 95\% confidence intervals shown in brackets.

\begin{table*}[t]
\caption{Effects of training data selection and evaluation scope on classification performance (BA, \%).}
\label{tab:filter-results}
\centering
\begin{tabular}{llrrl}
\hline
Model & Training set & Filtered-test BA (\%) & FULL-reference BA (\%) & $\Delta$BA (pp), 95\% CI \\
\hline
XGBoost & RULE & $82.05\pm0.32$ & $39.18\pm0.22$ & 42.87 [41.25, 44.66] \\
 & FULL & $77.99\pm0.57$ & $72.66\pm0.49$ & 5.33 [3.46, 7.38] \\
 & RANDOM & $56.11\pm0.37$ & $54.75\pm0.31$ & 1.36 [0.07, 2.71] \\
\midrule

YaTC & RULE & $44.45\pm2.15$ & $29.68\pm0.77$ & 14.77 [13.55, 15.92] \\
 & FULL & $30.46\pm0.45$ & $26.75\pm0.49$ & 3.71 [2.76, 4.58] \\
 & RANDOM & $32.66\pm0.74$ & $28.96\pm0.87$ & 3.69 [3.04, 4.38] \\
\midrule

ET-BERT & RULE & $39.17\pm0.63$ & $26.18\pm0.36$ & 12.99 [12.27, 13.68] \\
 & FULL & $26.23\pm1.36$ & $23.19\pm0.73$ & 3.04 [2.42, 3.63] \\
 & RANDOM & $26.23\pm2.27$ & $24.37\pm2.41$ & 1.86 [1.36, 2.36] \\
\hline
\end{tabular}
\end{table*}

\textbf{Changing evaluation scope alone exposes recognition failures hidden by the filtered benchmark.} All three RULE-trained models lose BA when excluded target flows are restored to scoring, with decreases of 12.99--42.87 percentage points. The full-reference test set includes additional connections belonging to the same website targets, which were excluded because they failed to match the SNI rule. The higher filtered scores therefore describe success on the rule-matched subset rather than recognition of all target flows.

\textbf{Filtered benchmarks can substantially overestimate a model's ability to perform the intended classification task.} XGBoost trained on RULE achieves 82.05\% BA when evaluated on SNI-matched flows. However, its BA drops to 39.18\% when evaluated on all independently attributed target flows. A researcher relying on the filtered benchmark may therefore consider the model sufficiently accurate for practical use, even though its performance on all independently attributed target flows admitted to FULL is much lower. The model remains unchanged; the benchmark simply excludes many target flows from evaluation.

The evaluation gap is distributed across website classes.
For XGBoost trained on RULE, recall decreases for 47 of
the 61 websites when excluded target flows are restored.
The five largest contributors account for only 7.30 of
the 42.87 percentage-point BA decrease. Decomposing
the gap by the requests carried on excluded flows shows
that connections carrying only cross-registrable-domain
requests contribute 38.89 percentage points (90.72\%)
of the total difference.

The RANDOM control helps determine whether the results are caused simply by fewer training samples or differences in class distribution. RANDOM and RULE contain the same number of flows from each website and collection run, but RANDOM selects flows without using SNI. For XGBoost, RULE performs better on the SNI-filtered test set, while RANDOM performs better on the complete target test set. This difference shows that the SNI rule affects which flows are selected, beyond reducing their number.

\takeaway{6}{When training and testing use the same restrictive filtering rule, models can achieve high benchmark scores while performing poorly on the full reference set of target flows. Such scores can lead researchers to overestimate the models' ability to perform the intended classification task.}

\guideline{5}{Dataset builders should provide independently attributed test data that includes target flows excluded by labeling rules. Downstream researchers should evaluate models on such data when available. If only filtered test data is available, researchers should clearly state the evaluation scope and avoid using benchmark scores alone to claim that a model is suitable for the intended classification task.}

\subsection{Auditability (RQ5)}
\label{sec:auditability-evaluation}

\textbf{Dataset releases do not always provide enough evidence to verify their labels and filtering decisions\cite{liu2022nidserrors}.} We examine whether researchers can recover the evidence used for labeling, link that evidence to individual flows\cite{buneman2001why,cui2000lineage}, and identify target flows excluded during dataset construction. We use CSTNET, NUDT, MIRAGE, AppClassNet, and \vt{} to examine these aspects.

\subsubsection{Preserving Labeling Fields}
\label{sec:label-fields}

\textbf{Releasing flow labels does not guarantee that researchers can verify how they were assigned.} CSTNET-TLS1.3 uses SNI for labeling, but we can recover ClientHello and a reliable SNI parsing path from only 4.98\% of its released flows. For the remaining flows, the released packets alone do not provide enough evidence to check the SNI-based labels. Independent verification requires the original identifier values and labeling rules to be preserved and linked to each flow.

Researchers may truncate packets or mask SNI to prevent information leakage into classification models. These processing steps need not remove the original labeling evidence from the dataset release. Keeping the labeling evidence separately allows researchers to train on processed network inputs while still checking how the labels were obtained.

\subsubsection{Verification Through Independent Evidence}
\label{sec:independent-records}

\textbf{Endpoint evidence can verify flow labels even when network identifiers are unavailable.} Using NUDT's application logs, connection records, and timestamps, we recover unambiguous application attribution for 68.46\% of processed flows~\cite{nudt2024}. Among the attributed TLS flows, 7.43\% have no recoverable SNI. Their application labels can still be verified through the endpoint records. This shows that network identifiers are not necessary when independent evidence establishes which application generated each connection.

Public datasets provide different levels of verification. MIRAGE marks whether each label comes from exact record matching or majority-based assignment, allowing researchers to identify the labeling method~\cite{mirage2019}. NUDT goes further by releasing application--connection records that allow individual flow labels to be reconstructed. In contrast, AppClassNet uses a proprietary DPI system to assign labels~\cite{appclassnet2022}. Without access to that system or its original decisions, researchers cannot independently reproduce its labeling results. Thus, knowing how labels were generated and being able to verify individual labels are different capabilities\cite{bujlow2015dpi}.

\textbf{Verifying which target flows were excluded requires evidence beyond the filtering rule itself\cite{dodge2021c4,grafberger2022}.} An SNI mapping can explain why a connection was assigned a particular service label, but it cannot identify other connections belonging to that service that failed the matching rule. CESNET-TLS22 acknowledges that its SNI--service mappings do not cover all relevant domains~\cite{cesnettls2023}. In \vt{}, independent visit--request--connection records establish which flows belong to each website visit. Together with the SNI-based filtering records, they allow researchers to identify both retained and excluded target flows.

\guideline{6}{Dataset builders should preserve attribution evidence and document how it supports each flow's label. They should also preserve evidence of sample selection, including which target flows were excluded. Derived datasets should retain links to their original samples and labels. Sensitive evidence may require controlled access, but independent verification should remain possible.}

\section{Discussion}
\label{sec:discussion}

\subsection{Benchmark Validity and Task Definition}

An ETC benchmark is valid when its labels and evaluated flows match the task it claims to measure\cite{bowmandahl2021,bean2025construct,cronbach1955construct}. SNI-based selection is valid when the benchmark
explicitly targets SNI-identifiable connections.
It becomes overstrict when scores on this subset
are used to assess recognition of all flows
belonging to an application or website visit.

Correctness and representativeness concern different failures:
a flow may receive an incorrect label, or a genuine target
flow may be absent from evaluation. Even when both
requirements are satisfied, flows from different targets may
have identical inputs under a given representation.
Excluding such flows would change the classification task
rather than correct a dataset defect. Auditability enables
researchers to distinguish problems of attribution and
coverage from limits of what a classifier can observe.

\subsection{Implications for Dataset Reuse and Method Evaluation}

Dataset reuse can carry original labeling assumptions into
new classification tasks. Flows may be segmented, combined
with other datasets, or assigned new classes, but these
operations cannot recover attribution evidence missing
from the original collection. When the prediction target
changes, researchers must check whether the inherited
labels remain supported. Preserving links among original
samples, labels, and processing
steps~\cite{iso5259part3,chapman2024} makes derived
datasets traceable and allows assumptions to be
reexamined~\cite{sambasivan2021}.

Progress in classification methods differs from progress
in benchmark quality. A higher score on a fixed benchmark
demonstrates better performance under its existing labels
and evaluation rules~\cite{northcutt2021}, not necessarily
better performance on the intended task. Correcting labels
or restoring excluded target flows may reduce model scores
while making evaluation more faithful to that task.
Lower scores can therefore accompany an improved benchmark.

Comparisons across studies require the same distinction.
Studies using the same dataset name and application classes
may apply different filtering rules or prediction targets,
making their scores
incomparable~\cite{holland2026pcapml,recht2019,engstrom2020bias}.
Even with identical evaluation procedures, gains on a
restricted benchmark establish progress only on the task
it measures~\cite{torralba2011,zech2018,mccoy2019heuristics,ribeiro2020checklist}.
Claims about broader classification tasks need
corresponding evidence~\cite{juarez2014,oakdenrayner2020,sommer2010closed}.
Reliable assessments of ETC progress require trustworthy
evaluation data of dataset construction
and reuse~\cite{bender2018datastatements,akhtar2024croissant}.

\subsection{Open Research Questions}
\label{sec:open-questions}

Our findings also raise questions about how reliable ETC datasets can be constructed at scale and how benchmark classification tasks should be defined.

\textbf{Reliable labeling of diverse real-world traffic.}
Automated traffic labeling has followed several approaches.
Context-based collection assigns labels according to the
application or activity executed during a capture. Endpoint-
assisted systems associate applications with flows through
process, socket, or connection information, as demonstrated
by GT and A3C~\cite{dusi2011,a3c2022}. Browser-assisted
methods further use browser observations to associate traffic
with browsing activities~\cite{zaki2021grano}. Network-side
systems such as DN-Hunter and WHAT infer domain or
website-related traffic associations from DNS, addresses,
and temporal relationships~\cite{bermudez2012dns,what2016}.
These approaches differ in what their evidence establishes:
the originating application, the remote domain, and the
website visit are distinct classification targets.

The remaining challenge is to obtain diverse, naturally
occurring traffic while maintaining reliable attribution.
Endpoint-assisted collection provides stronger attribution
evidence but may not cover the diversity of operational
networks. Traffic-only observations can capture naturally
occurring traffic across different environments, but lack
direct endpoint evidence. One direction is to combine
network-side observations into traceable attribution chains,
while explicitly distinguishing supported and uncertain
labels and measuring both label correctness and target-flow
coverage. Flows without sufficient evidence need not be
assigned labels; they can still support self-supervised
pretraining~\cite{etbert2022}. Evaluation of the intended
classification task, however, requires independently
attributed reference data.

\textbf{Choosing the unit of classification.}
Flows provide a practical unit for online monitoring because
their state and features can be updated as packets arrive.
However, the unit used to process traffic need not match
the classification target. A website visit may involve
multiple flows, while a connection may carry multiple
requests. The open question is when an individual flow
contains sufficient information to identify the target and
when classification requires context from related flows or
a larger session. Benchmarks should support comparisons
that account for classification performance, decision
latency, and processing cost~\cite{ethayarajh2020utility}.
\section{Conclusion}
\label{sec:conclusion}
We trace how 13 ETC datasets turn evidence into per-flow labels and evaluation sets. Coarse inheritance can hide losses in application recognition behind higher accuracy. Overstrict filtering can exclude target flows and reverse the ranking of models trained on different training sets.

In conclusion, reliable ETC benchmarking therefore requires \textbf{correctness, representativeness, and auditability (CRA)}. Labels need evidence at the prediction target's granularity, and evaluation must cover the target flows. Preserving evidence for both lets researchers judge whether higher benchmark scores reflect better performance on the intended classification task.
\cleardoublepage

\nocite{iso5259,mls,lowquality2024}
\bibliographystyle{IEEEtran}
\bibliography{ref}

\appendices

\begin{table*}[t]
\centering
\caption{Definitions of key terms used in this SoK.}
\label{tab:terminology}
\begin{tabular}{@{}L{0.16\linewidth}L{\dimexpr0.84\linewidth-2\tabcolsep\relax}@{}}
\hline
\textbf{Term} & \textbf{Definition} \\
\hline

Dataset &
A collection of network traffic samples and assigned labels,
together with any accompanying construction materials. \\

Benchmark &
An evaluation task defined using a dataset, including its classification
target, evaluation scope, and scoring procedure. A dataset may support
multiple benchmarks. \\

Classification target &
The identity or property a classifier is intended to predict, such as
an application, website visit, service, or activity. \\

Target flow &
A flow belonging to the declared classification target, regardless
of whether a labeling or filtering rule can identify it. \\

\hline

Attribution evidence &
Information supporting the association between an individual flow
and its claimed classification target. \\

Evidence granularity &
The level at which evidence applies, such as a collection run,
activity, process, request, or connection. \\

Record &
A concrete capture or log, such as PCAP, HAR, or NetLog, that may
contain attribution evidence. Its existence alone does not establish
per-flow attribution. \\

Independent attribution &
Establishing flow--target associations using evidence independent
of the network-identifier rule being evaluated. \\

\hline

Context inheritance &
Assigning the label of a collection run or activity to its network
samples based on their collection context. \\

Coarse inheritance &
Assigning collection-level labels to individual flows without
corresponding per-flow attribution evidence. \\

Majority assignment &
The MIRAGE experimental procedure that assigns each flow the
most common application label among other matched flows in its capture. \\

Record association &
Assigning flow labels by matching endpoint or connection records
to network flows. \\

\hline

Sample selection &
Rules determining which collected flows are retained for dataset
release or downstream evaluation. \\

Protocol filtering &
Selecting flows according to protocol types or related conditions,
thereby changing the traffic scope of the classification task. \\

Overstrict filtering &
Excluding target flows that fail network-identifier matching rules
used for labeling or sample selection. \\

Evaluation scope &
The set of flows contributing to a reported benchmark score, such
as a filtered subset or an independently attributed reference set. \\

\hline

Input conflict &
Flows with different labels but identical model inputs under a fixed
representation. Such conflicts do not necessarily indicate labeling errors. \\

Empirical BA ceiling &
The maximum balanced accuracy achievable on a fixed labeled
sample set and input representation when identical inputs must receive
identical predictions. \\

\hline

Correctness &
The requirement that each flow's label correspond to the
classification target it actually belongs to. \\

Representativeness &
The requirement that evaluation cover the target traffic specified
by the intended classification task. \\

Auditability &
The requirement that sufficient evidence be preserved to
independently verify flow labels and sample-selection decisions. \\

\hline
\end{tabular}
\end{table*}

\section{Ethics Considerations}
\label{app:ethics}

This study examines benchmark validity using public ETC datasets and traffic from controlled website visits. It aims to improve the reliability of method comparisons and reduce misleading conclusions about recognition capabilities. Affected parties include people represented in existing datasets, operators of visited websites, dataset maintainers, and users of subsequent research.

Traffic captures and endpoint records can reveal browsing behavior and contain sensitive identifiers; public availability alone does not remove these concerns. Our audit relies on associations among samples, labels, and collection records and does not require identifying individual users. Thus, preserving attribution evidence does not imply unrestricted publication of raw endpoint records: stable sample identifiers and authorized access to sensitive records can reconcile per-flow verification with confidentiality. Controlled visits to live websites also consume third-party resources, a relevant consideration in collecting reference traffic.

Improved traffic classification has defensive applications and potential privacy-invasive uses. This paper addresses the validity of benchmark supervision and evaluation. We distinguish missing attribution evidence from demonstrated labeling errors, and genuine input ambiguity from defects in label construction. These distinctions support evidence-based dataset assessment without treating every unresolved attribution relationship as an incorrect label.

\section{Generative AI Usage Considerations}
\label{app:genai}

Generative AI was used for English translation, language editing, terminology harmonization, LaTeX preparation, and drafting of disclosure text. It reduced the editorial and formatting workload. The authors are responsible for the manuscript's claims, citations, numerical results, and disclosures.

\section{Terminology}
\label{app:terminology}

Table~\ref{tab:terminology} defines the key terms used throughout this SoK and clarifies distinctions between closely related concepts.

\section{VisitTrace}
\label{app:visittrace-attribution}

VisitTrace can be accessed via the following anonymous link: \url{https://figshare.com/s/949a4079e2ff8a6ab2da.}

\subsection{Attribution and Inclusion}
\textbf{Collection and attribution coverage.}
The common 61-class \vt{} release comprises 2,223 website
visits, 268,684 HTTP Archive (HAR) requests, 306,357
Chrome DevTools Protocol (CDP) request observations,
105,376 visit-scoped NetLog sockets, and 619,094
bidirectional TCP flows captured in PCAP files.
An additional 164,048 UDP endpoint groups were counted
separately and are not included in the TCP-flow analysis.

Table~\ref{tab:visittrace-funnel} summarizes the attribution
process. Of the 306,357 CDP request observations, 268,116
(87.52\%) have verified native associations. Among the
105,376 NetLog sockets, 96,223 (91.31\%) can be matched
to distinct TCP flows in the packet captures. Of these
matched flows, 37,284 (38.75\%) have at least one verified
native browser-request association. After applying the
complete-flow attribution requirements, the final FULL
set contains 36,143 flows.

\begin{table}[t]
\caption{Collection and attribution coverage of the common
61-class VisitTrace release. Rates use the denominators
shown in the third column.}
\label{tab:visittrace-funnel}
\centering
\begin{tabular}{@{}>{\raggedright\arraybackslash}p{0.4\columnwidth}rrr@{}}\hline
\textbf{Stage} &
\textbf{Count} &
\textbf{Denominator} &
\textbf{Rate} \\
\hline
Website visits
    & 2,223 & -- & -- \\
HAR request entries
    & 268,684 & -- & -- \\
CDP request observations
    & 306,357 & -- & -- \\
NetLog sockets
    & 105,376 & -- & -- \\
Captured bidirectional TCP flows
    & 619,094 & -- & -- \\
\hline
CDP observations with verified native associations
    & 268,116 & 306,357 & 87.52\% \\
Sockets matched to PCAP TCP flows
    & 96,223 & 105,376 & 91.31\% \\
Matched flows with verified native requests
    & 37,284 & 96,223 & 38.75\% \\
Final FULL flows among matched browser flows
    & 36,143 & 96,223 & 37.56\% \\
Final FULL flows among all captured TCP flows
    & 36,143 & 619,094 & 5.84\% \\
\hline
\end{tabular}
\end{table}

\textbf{Request association and ambiguity.}
HAR and CDP observations provide complementary information
about browser requests. Matching the 268,684 HAR entries
against CDP observations from the same visit using HTTP
method and exact URL produces 223,465 entries (83.17\%)
with a unique candidate, 45,216 (16.83\%) with multiple
candidates, and three without a candidate. The complete
association procedure additionally uses request, response,
and socket evidence. A unique method--URL match alone
does not establish the final flow attribution.

Of the 306,357 CDP observations, 38,241 lack a verified
native association. These comprise 19,943 duplicate
observation sessions, 13,078 observations without a direct
socket identifier, and 5,220 requests handled through
caching or a local Service Worker. These cases are not
treated as independently attributed network flows solely
because their requests were observed.

\textbf{Socket-to-flow association.}
NetLog socket information is matched against the endpoints
and lifetimes of captured TCP flows. Of the 105,376 sockets,
9,153 lack complete endpoint information. All 96,223
sockets with complete endpoints have corresponding PCAP
candidates. For 96,221 sockets, the endpoints identify a
unique flow. Two sockets have multiple candidates because
the same endpoints were reused at different times. Comparing
the TCP connection-establishment times with packet
timestamps resolves both cases, leaving no unresolved
socket-to-PCAP ambiguity among sockets with complete
endpoints.

Connection failures are distinguished from successful
associations. Among 102,298 sockets with an observed
TCP\_CONNECT start, 707 explicitly report connection
errors, 5,366 have no observed completion, and two report
address-retrieval errors at completion. The remaining
96,223 successfully establish complete endpoint information
and are matched to captured flows.

\textbf{Complete-flow inclusion.}
A browser socket association alone is insufficient to assign
a website-visit label. A flow is admitted to FULL only when
the available attribution evidence supports assigning the
complete flow to one website visit. Connections without
verified native browser-request associations are not
automatically assigned the visit label. Additional evidence
replay can recover attribution for some such connections;
otherwise, they remain outside FULL.

Among the 37,284 flows with verified native request
associations, 1,197 do not satisfy the final inclusion
conditions because request attribution remains unresolved.
Evidence replay additionally admits 56 flows without native
verified request associations. Therefore, the final count is 36,143.

Equivalently, the FULL set contains 35,794 flows admitted
through native attribution and 349 flows admitted using
recovered evidence. Of the recovered flows, 293 were
already included among the 37,284 flows with native
request associations; the other 56 were added through
recovery.

Across the published flows, 232,209 distinct CDP request
keys can be traced through the association evidence.
Among 8,926 recovered send events, 33 can be associated
only with candidate-request sets rather than a unique
browser request. Their available evidence nevertheless
establishes the website-visit attribution of the corresponding
flows. Thus, uncertainty about an individual request does
not necessarily imply uncertainty about the visit label
of the complete flow.

\textbf{Connection reuse within visits.}
Persistent TCP connections are retained within a website
visit. Among the 35,794 flows admitted through native
request associations, 18,399 (51.40\%) carry at least two
distinct browser requests. A single flow carries as many
as 727 requests. Of these multi-request flows, 15,281 use
HTTP/2 and 3,118 use HTTP/1.1. Among the 349 flows
admitted through recovered attribution evidence, 303
contain multiple observed send events.

Multiple requests sharing a connection do not create
additional flow samples. Each admitted TCP flow contributes
exactly one sample to \vt{} and to the H5/H20 input-conflict
analyses, regardless of the number of requests or send
events associated with it.

\textbf{Attribution of shared connections.}
A connection can carry requests to different remote origins
within the same website visit. Among the natively attributed
flows, 1,571 contain requests to multiple origins, with up
to ten origins associated with a single flow. Each request
is checked against its website-visit attribution, browser
socket, and corresponding PCAP flow.

Different remote origins do not necessarily imply different
website-visit labels. For example, requests to a website's
own servers and third-party resource servers may all support
the same visit. However, a connection carrying both
page-related requests and browser-component traffic cannot
be attributed exclusively to the page. Accordingly, 119
flows with mixed page and component usage are excluded
from FULL.

\textbf{Connection boundaries across visits.}
Each of the 2,223 published visits launches an independent
Chromium process with a temporary browser profile.
Collection intervals do not overlap, and packet capture
continues until the corresponding browser process exits.
All 36,143 published FULL flows begin with a TCP SYN
that establishes a new connection.

Across visits, 80 endpoint pairs recur, involving 160 TCP
flows. For each recurring endpoint pair, the client initial
TCP sequence numbers differ between visits, confirming
separate connection establishments. No published browser
TCP connection is observed to be reused across visits.

This design permits the study of persistent connections and
request multiplexing within individual website visits. It
does not measure connection reuse across different website
visits during continuous browsing in a shared browser
session.

\textbf{Disposition of excluded flows.}
The FULL inclusion requirements exclude 582,951 of the
619,094 captured TCP flows. Table~\ref{tab:visittrace-exclusion}
summarizes their disposition. The largest group contains
flows whose attribution remains unresolved or unknown.
Other flows have been identified as background traffic,
preconnections without verified page requests, mixed-use
connections, or protocol-assistance traffic.

\begin{table}[t]
\caption{Disposition of captured TCP flows not admitted
to the VisitTrace FULL set.}
\label{tab:visittrace-exclusion}
\centering
\begin{tabularx}{\linewidth}{Xr}   
\hline
\textbf{Disposition} & \textbf{Flows} \\
\hline
Unresolved or unattributed
    & 487,046 \\
Connections or preconnections without verified page requests
    & 37,788 \\
Confirmed background traffic
    & 32,278 \\
Candidate background traffic
    & 25,491 \\
Confirmed page-related flows outside the sample window
    & 127 \\
Mixed page and browser-component traffic
    & 119 \\
Browser protocol-assistance traffic
    & 102 \\
\hline
Total excluded TCP flows
    & 582,951 \\
\hline
\end{tabularx}
\end{table}

\textbf{Interpretation of attribution coverage.}
The 36,143 admitted flows represent 5.84\% of all captured
TCP flows and 37.56\% of TCP flows matched to browser
sockets. These percentages describe the coverage achieved
under the FULL attribution requirements. They are not
estimates of the proportion of all genuine website-related
flows identified, because the excluded set includes both
known non-target traffic and flows whose attribution cannot
be established.

The SNI-based filtering experiment is conducted separately,
after independent flow attribution. Among the 36,143
admitted target flows spanning the same 61 website classes,
4,908 satisfy the target-domain SNI matching rule, yielding
a retention rate of 13.58\%. This measures the coverage of
the SNI rule within an independently attributed reference
set, rather than within all captured TCP traffic.

These corpus-specific measurements demonstrate how
network-identifier filtering can exclude independently
attributed target flows. They do not assume that other
ETC datasets exhibit the same retention rate or
classification-performance differences.

\subsection{Website-Level Filtering and Evaluation Effects}
\label{app:visittrace-filtering}

\textbf{Variation across websites.}
SNI-based filtering affects websites unevenly. Across the
common 61-class VisitTrace corpus, the rule retains
4,908 of 36,143 independently attributed target flows
(13.58\%). The mean retention rate across websites,
giving each website equal weight, is 43.34\%, with a
median of 34.55\%. Nevertheless, 38 of the 61 websites
lose more than half of their target flows.

The aggregate retention rate is influenced by websites
generating large numbers of flows. For example,
aljazeera.com retains only 71 of 12,355 target flows
(0.57\%), accounting for 39.33\% of all excluded flows.
Excluding this website from the calculation raises the
aggregate retention rate to 20.33\%. Thus, the exact
retention rate depends on the website distribution, but
substantial target-flow exclusion is not confined to a
single high-volume website.

\textbf{Request-domain relationships.}
To examine which connections are excluded, we compare
the hostnames of associated browser requests with the
top-level page visited by the user. Registrable domains
are determined using the Public Suffix List, including
its PRIVATE section. A flow is classified as same-domain
only, cross-domain only, or mixed according to the
domains of all its associated requests. Flows without
sufficient domain information remain undetermined.

Table~\ref{tab:visittrace-domain} reports SNI retention
for these mutually exclusive groups.

\begin{table}[t]
\caption{SNI retention by request-domain relationship
in the common 61-class VisitTrace corpus.}
\label{tab:visittrace-domain}
\centering
\begin{tabular}{lrrr}
\hline
\textbf{Request origins} &
\textbf{FULL} &
\textbf{Retained} &
\textbf{Rate} \\
\hline
Same-domain only  & 5,678  & 4,800 & 84.54\% \\
Cross-domain only & 30,377 & 72    & 0.24\% \\
Mixed             & 36     & 36    & 100\% \\
Undetermined      & 52     & 0     & 0\% \\
\hline
Total             & 36,143 & 4,908 & 13.58\% \\
\hline
\end{tabular}
\end{table}

Of the 31,235 excluded target flows, 30,305 (97.02\%)
carry only requests to registrable domains different from
that of the visited page. Furthermore, 31,200 excluded
flows have visible but nonmatching SNI; only 35 have
unavailable SNI. Thus, target-flow exclusion in this
corpus arises predominantly from domain mismatches,
rather than missing SNI.

The request-domain grouping and the SNI rule are not
identical. The rule matches frozen target domains and
their predefined aliases, rather than all domains sharing
the same registrable domain. Consequently, 72 flows
classified as cross-domain only are retained through
predefined aliases, while 878 same-domain-only flows
are excluded. Of the latter, 847 have SNI within the
visited page's registrable domain but do not satisfy
the frozen matching rule. Registrable-domain relationships
also describe resource locations, not necessarily
organizational ownership.

\textbf{Distribution of evaluation differences.}
We further examine the 42.87 percentage-point BA
difference reported for XGBoost trained on RULE
(Table~\ref{tab:filter-results}). The filtered
evaluation contains 1,610 target test flows, compared
with 11,669 flows in the complete reference evaluation.
Both evaluations use the same saved predictions from
three training seeds and the same 61 website classes.

Restoring excluded flows reduces recall for 47 websites;
the remaining 14 show no change, and none improve.
The median per-website recall decrease is 45.83 percentage
points. The five largest website-level contributors
account for 7.30 percentage points (17.04\%) of the
total BA decrease. Contributions from 17 websites are
needed to account for half of the decrease, showing
that the result is not driven by a few website classes.

To identify which excluded flows contribute to this
difference, let $r_c$ denote the mean recall on the
filtered test flows of website $c$, and let $n_c$
denote its number of complete-reference test flows.
For each excluded flow $i$, let $a_i$ be its mean
correctness across the three saved predictions, and
let $c(i)$ denote its website class. The difference
in BA, expressed in percentage points, can be written as

\begin{equation}
\Delta \mathrm{BA}
=
\frac{100}{61}
\sum_{i\in E}
\frac{r_{c(i)}-a_i}{n_{c(i)}},
\label{eq:visittrace-gap-decomposition}
\end{equation}

where $E$ is the set of excluded target test flows.
This decomposition keeps the models and predictions
fixed and attributes the score difference to the
excluded evaluation samples.

Among the excluded test flows, 9,765 carrying only
cross-domain requests contribute 38.89 percentage
points (90.72\%) of the total difference. Another
278 same-domain-only flows contribute 3.93 percentage
points, while 16 flows with undetermined domain
relationships contribute 0.05 percentage points.
These contributions reconstruct the reported 42.87
percentage-point difference. They identify where
the evaluation gap occurs, without attributing the
same effect magnitude to other datasets or models.

\textbf{Repeatability within the collection period.}
The three predefined collection rounds, corresponding
to the training, validation, and test partitions, retain
13.55\%, 13.40\%, and 13.80\% of target flows,
respectively. Collection took place between October 8
and 23, 2026 (UTC). For 743 configurations with
observations on multiple days, covering 60 websites,
the website-weighted retention rates on the first
and last observed days are 42.47\% and 42.56\%.
The median absolute per-website change is 0.19
percentage points. These results indicate that the
aggregate coverage difference persists across the
collection rounds, although individual websites
can vary over time.

\end{document}